\documentclass[amssymb,amsfonts,pra,aps,floatfix,showpacs]{revtex4}
\usepackage[dvips]{graphicx}
\usepackage{varioref,epsfig}
\begin{document}
\title{\bf $1/N$-expansions in non-relativistic quantum mechanics\\}
\author{Niels Emil Jannik Bjerrum-Bohr}
\address{University of Copenhagen\\ The Niels Bohr
Institute\\
Blegdamsvej 17, DK-2100 Copenhagen\\
Denmark} \draft
\date{\today}
\begin{abstract}
An extensive number of numerical computations of
energy 1/$N$ series using a recursive Taylor series method are
presented in this paper.
The series are computed to a high order of approximation
and their behaviour on increasing the order of approximation is examined.
\end{abstract}
 \maketitle

\section{Introduction}
In chemistry and atomic physics, $1/N$ (or $1/D$) perturbation theory is
very powerful in calculating energy eigenstates for many complicated
systems\cite{alin,dtp,lod,loo}.
The method is especially valuable when the applicability of Hartree-Fock methods
is limited. In this study the behaviour of $1/N$ expansions is investigated at a
very high order of approximation. For the sake of simplicity only radial potentials
with spherical symmetry will be considered.
The $N$-dimensional non-relativistic Schr\"odinger equation for such potentials
may be written as
\begin{equation}
\left  [-\frac{1}{2}\frac{d^2}{d r^2} -
\frac{1}{2}\frac{N-1}{r}\frac{d}{d r}
+\frac{l(l + N - 2)}{2r^2} + V(r) \right]\phi(r) = E \phi(r)
\end{equation}
where $N$ is the number of spatial dimensions \cite{louck}.
A brief review of this equation is presented in appendix \ref{appendix2}.

The concept underlying $1/N$ expansion methods is the use of perturbation theory
on the expandable parameter $1/N$.
A standard problem in perturbation theory would be a solvable Schr\"{o}dinger
equation plus a small perturbation, expressed as
\begin{equation} H = H_0 + gV \end{equation}
where $H_0$ is the solvable Hamiltonian and $gV$ is the small perturbation. In
this notation $g$ is a real expansion coefficient.
To optain an approximation to the eigenstates of the perturbed
Schr\"{o}dinger equation one inserts a series for the
wave function in powers of $g$ as
\begin{equation}
u = u^{(0)} + gu^{(1)} + g^2u^{(2)} + ...
\end{equation}
and a series in energy as
\begin{equation}
E = E^{(0)} + gE^{(1)} + g^2E^{(2)} + ...
\end{equation}
Finally, these series are solved for the unknown functions $u^{(\nu)}$ and the
energy coefficients $E^{(\nu)}$.

For the standard problem this procedure provides an efficient way
to obtain an estimate of the eigenstates and the energy of the
physical system. Widespread application of this method is,
however, limited since only problems with an expandable
coefficient $g$ can be solved. The dimensional dependency of the
Schr\"odinger equation provides us with a hidden expandable
variable of the potential. Roughly speaking, enlarging the
dimensions of a physical system turns the system ``classical'',
and in the limit of large dimensionality a particle in a potential
is fixed at the minimum; in some sense, $\hbar \rightarrow 0$. A
classical analog could be a damped pendulum. A differential
equation describing this is e.g.:
\begin{equation}\nonumber
mr\ddot{\theta}+b\dot{\theta}+ mg\theta = 0
\end{equation}
where $mr\ddot{\theta} \ll b\dot{\theta}$ (assuming small angles).
In the case of a very heavily damped pendulum it is safe to
disregard the $\ddot{\theta}mr$ term. This implies that the
second-order differential equation reduces to the first-order
differential equation
\begin{equation}\nonumber
b\dot{\theta}+ mg\theta = 0
\end{equation}
Solutions of a first-order differential equation are unable to
oscillate.
The pendulum thus seeks the fixed point of the potential.\\

An $N$-dimensional particle, where $N$ is very large, behaves somewhat like
a pendulum in a highly viscous fluid.
Its ``damping term'' (the first order differential term) in the Schr\"odinger
equation is large, and hence its de Broglie wavelength becomes small.
To leading order in $1/N$, the particle may be addressed by quasi-classical
concepts.
Accordingly, the fixed point of the potential provides a starting point for a
power series expansion in $1/N$ for the energy eigenvalue and the wave
function.

Most $1/N$-series provide accurate results even when only the first few terms
are summed; it is thus more interesting to examine the behaviour of the series
as the order of approximation increases.
An interesting aspect of the $1/N$-method is that it is perturbatively
exact in the sense that there are no explicit approximations in the method.
Most potentials lead to apparently asymptotic series at
large orders of approximation.
The approach here is to calculate the $1/N$ series to a significant
order (30 - 100 terms) for miscellaneous potentials and then explore the
behaviour of the series. This is accomplished by employing a computer algebraic
program. Round-off errors are also taken account of in the calculations.

\section{Theory}
\subsection*{How to obtain the $1/N$-series}
In order to obtain $1/N$-series for a given potential there are several ways
to proceed. One procedure is to expand the wave function, the energy, and the
potential in Taylor series \cite{stse}.
This approach makes it easy to make an efficient computer algorithm.
Before the Taylor series expansions are made, it is preferable to redefine the
equations slightly.
First, the $N$-dimensional Schr\"odinger equation is transformed into
\begin{equation}
\left  (-\frac{1}{2} \frac{d^2}{d r^2} +
k^2\left  [\frac{(1-\frac{1}{k})(1-\frac{3}{k})}{8r^2}\right ]
 + {\hat V(r)} \right   ) \psi(r) = E' \psi(r)
\end{equation}
by redefining $\psi(r) = r^{\frac{(N-1)}{2}}\phi(r)$.
Now, $k$ is defined by $k = N + 2l$. The energy eigenvalue is denoted by E'.\\
The position $r$ is then redefined in terms of $k$, $r = \sqrt{k}\rho$
or $yr = \rho$ where $y$ is $\frac{1}{\sqrt{k}}$.
The Schr\"odinger equation is subsequently recast into to a
differential equation in $\rho$, as
\begin{equation}
\left  (-\frac{1}{2} \frac{d^2}{d \rho^2} +
k^2 \left  [\frac{(1-\frac{1}{k})(1-\frac{3}{k})}{8\rho^2}
 + \frac{\hat V(\rho \sqrt{k})}{k} \right ] \right ) \psi(\rho) = E'k\psi(\rho)
\end{equation}
If now $\hat V(r) = kV(\rho)$, then
\begin{equation}
\left  (-\frac{1}{2} \frac{d^2}{d \rho^2} +
k^2 \left  [\frac{(1-\frac{1}{k})(1-\frac{3}{k})}{8\rho^2}
 + V(\rho)\right   ] \right   ) \psi(\rho) = E'k\psi(\rho) \label{Eqn:rhos}
\end{equation}
In the large-$N$ limit one has an effective potential like
\begin{equation}
k \left  [\frac{1}{8\rho^2} +
V(\rho) \right ]
\end{equation}
Its minimum is the energy $kE^{(-2)}$ which matches the energy at the minimum
$\rho_0$.\\
To ease the further derivations it is preferable to rescale distance as
$x = \sqrt{k}(\rho - \rho_0)$
and define $\psi(\rho) = e^{U(x)}$ and
\begin{equation}
V_{\mathrm{eff}}(x) = \frac{1}{8}\rho(x)^{-2} + V(\rho) -
E^{(-2)}
\end{equation}
Introducing $\psi$ in eq. (\ref{Eqn:rhos}) leads to a differential
equation for $U(x)$. \\
\begin{equation}
-\frac{1}{2}\left  [ \ddot{U}(x) + \dot{U}(x)^2 \right   ] +
kV_{\mathrm{eff}}(x)
+ \left (-\frac{1}{2} + \frac{3}{8}k^{-1}\right )\rho(x)^{-2} = E' - E^{(-2)}k
= E
\end{equation}
or
\begin{equation}
\ddot{U}(x) + \dot{U}(x)^2 -
2W(x) + 2E = 0\label{Eqn:req}
\end{equation}
where $W(x) = kV_{\mathrm{eff}}(x) + \left [ \frac{-\frac{1}{2} +
\frac{3}{8}k^{-1}}{\rho(x)^2}\right ]$.
Now the actual Taylor series expansion may begin.
Extending $\dot{U}(x)$, $E$, $W(x)$ in
Taylor series in $x$ and $y$ one obtains.
\begin{eqnarray}
U(x) = \sum_{n=0}^\infty \sum_{m=0}^{n+1} \left
[\frac{D_m^nx^{2m}}{2m}
\right   ]y^{2n}
+\sum_{n=0}^{\infty} \sum_{m=0}^{n+1}\left  [
\frac{C_m^nx^{2m+1}}{2m+1}\right   ] y^{2n+1}
\end{eqnarray}
\begin{eqnarray}
E= \sum_{n=0}^{\infty}{E^{(n-1)}}{y^{2n}}
\end{eqnarray}
Coefficients of the type $D_0^n$ are defined to be zero.
The series for the potential is extraordinary. For $W(x)$ it turns out that
for any power of $x$, say $k$ $(k > 1)$, one has solely terms of
$y$ in powers ${k-2}$, ${k}$ and ${k+2}$.

\subsection*{The recursion formulas}
Using the Taylor series technique, series expansions for
the energy and the wave function have to be found by recursion.
The easiest way to do this is to find universal recursion
formulas \cite{stse} for the required
coefficients and then use a computer to recursively calculate
each coefficient one by one to a preferred number of terms.
For the ground state, the recursion relations are established by the requirement that
\begin{equation}
\ddot{U}(x) + \dot{U}(x)^2 - 2W(x) + 2E=
0
\end{equation}
or
\begin{eqnarray}
\sum_{n=0}^\infty \sum_{m=0}^{n} \left  [(2m+1){D_{m+1}^n}
x^{2m}\right   ]y^{2n}
+\sum_{n=0}^{\infty} \sum_{m=0}^{n+1}\left  [(2m)
{C_m^nx^{2m-1}}\right   ] y^{2n+1} \nonumber \\
+\left [\sum_{n=0}^\infty \sum_{m=0}^{n+1} \left [{D_m^n}
x^{2m-1}\right ]y^{2n}
+\sum_{n=0}^{\infty} \sum_{m=0}^{n+1}\left  [
{C_m^nx^{2m}}\right   ] y^{2n+1}\right ]^2 +
2\sum_{n=0}^{\infty}{E^{(n-1)}}{y^{2n}}\nonumber \\
-2\left [ W_0^0 + W_2^0x^2 + \left  (W_1^1x + W_3^1x^3\right   )y +
\sum_{n=1}^{\infty}\left  (W_{2n-2}^{2n}x^{2n-2} +
W_{2n}^{2n}x^{2n} + W_{2n+2}^{2n}x^{n+2} \right   )y^{2n}\right .
\nonumber \\ \left .+ \sum_{n=1}^{\infty}\left  (W_{2n-1}^{2n+1}x^{2n-1} +
W_{2n+1}^{2n+1}x^{2n+1} + W_{2n+3}^{2n+1}x^{2n+3} \right   )y^{2n+1}
\right
] = 0
\end{eqnarray}
From these equations it is apparent that
\begin{equation}
D_1^0 = -\sqrt{2W_2^0}
\end{equation}
and
\begin{equation}
C_{n+2}^{n} = D_{n+2}^{n} = 0
\end{equation}
Likewise, for the coefficients $D_m^n$ it follows that
\begin{eqnarray}
D_{m}^{n} = -\frac{1}{2D_1^0}{\left  [-2W_{2m}^{2n} + (2m+1)D_{m+1}^{n}
+\sum_{i=1}^{n-1}\sum_{j=1}^{i+1}D_j^iD_{m+1-j}^{n-i}
+\sum_{i=0}^{n-1}\sum_{j=0}^{i+1}C_j^iC_{m-j}^{n-i-1} \right]}
\end{eqnarray}
For the coefficients $C_m^n$ one obtains
\begin{eqnarray}
C_{m}^{n} = -\frac{1}{2D_1^0}{\left  [-2W_{2m+1}^{2n+1} + 2(m+1)C_{m+1}^{n}
+2\sum_{i=1}^{n}\sum_{j=1}^{i+1}D_j^iC_{m+1-j}^{n-i} \right ]}
\end{eqnarray}
And for the energy one obtains, after some algebra,
\begin{eqnarray}
E^{(n-1)} = \frac{1}{2}\left  [-D_1^n + 2W_{0}^{2n} -
\sum_{i=0}^{n-1}C_0^{i}C_0^{n-i-1}\right ]
\end{eqnarray}
All coefficients of the type $C_j^n$, where $j$ is negative, or $D_j^n$,
where $j$ is zero or negative, are assumed to be zero in the summation.

Obvious extensions of this technique to include excited states are possible.
We can always write
\begin{equation}
\psi = \left (xy - A(y)\right )e^{U(x)}
\end{equation}
for the first excited state, and
\begin{equation}
\psi = \left (x^2y^2 + xyC(y) +B(y)\right )e^{U(x)}
\end{equation}
for the second excited state, etc.
Here, $A(y)$, $B(y)$ and $C(y)$ are series expansions
in $y$ for the nodes in $\psi$.
\begin{equation}
A(y) = \sum_{n=1}^{\infty} a_n y^{2n} \ \ \ \
B(y) = \sum_{n=1}^{\infty} b_n y^{2n} \ \ \ \
C(y) = \sum_{n=1}^{\infty} c_n y^{2n}
\end{equation}
Subsequently, a differential equation for $U(x)$ is obtained for the first excited
state.
\begin{eqnarray}
\left (xy - A\right )\left (\ddot{U}(x) + \dot{U}(x)^2 -
2W(x) +
2E\right ) + 2y\dot{U}(x) = 0
\end{eqnarray}
Expressing
\begin{equation}
\ddot{U}(x) + \dot{U}(x)^2 - 2W(x) + 2E
\end{equation}
in a new Taylor series
\begin{equation}
{T}+ {S} = \sum_{n=0}^{\infty}\sum_{m=0}^{n}\left
[T_m^nx^{2m}\right ]y^{2n}
+ \sum_{n=0}^{\infty}\sum_{m=0}^{n}\left [ S_m^nx^{2m+1}\right ]y^{2n+1}
\end{equation}
the previously obtained universal recursion formulas for the ground
state are again useful.
The recursion formulas for ${T}$ and ${S}$ are arrived at using the relation
\begin{eqnarray}
\left (xy - A\right )\left ({T}+{S}\right ) +
2y\dot{U}(x) = 0
\end{eqnarray}
or
\begin{eqnarray}
\left [xy - \sum_{n=1}^{\infty} a_n y^{2n}\right ]
\left [\sum_{n=0}^{\infty}\sum_{m=0}^{n}\left [T_m^nx^{2m}\right ]y^{2n}
+ \sum_{n=0}^{\infty}\sum_{m=0}^{n}\left [ S_m^nx^{2m+1}\right ]y^{2n+1}
\right ]
\nonumber \\ +2y\left [\sum_{n=0}^\infty \sum_{m=0}^{n+1} \left [{D_m^n}
x^{2m-1}\right ]y^{2n}
+\sum_{n=0}^{\infty} \sum_{m=0}^{n+1}\left  [
{C_m^nx^{2m}}\right   ] y^{2n+1}\right ] = 0
\end{eqnarray}
One obtains
\begin{equation}
T_m^n = \sum_{k=1}^{n-m} a_kS^{n-k}_m - 2D_{m+1}^n
\end{equation}
and
\begin{equation}
S_m^n = \sum_{k=1}^{n-m} a_kT^{n-k+1}_{m+1} - 2C_{m+1}^n
\end{equation}
\begin{equation}
a_n = \frac{1}{T_0^0}\left [ {2C_0^{n-1} -
\sum_{k=1}^{n-1} a_kT^{n-k}_0}\right ]
\end{equation}
Once again
\begin{equation}
D_1^0 = -\sqrt{2W_2^0}, \ \ \ \ \ \ \ \ \ \ \ \
\ \ \ \ \ C_{n+2}^{n} = D_{n+2}^{n} = 0
\end{equation}
\begin{eqnarray}
D_{m}^{n} = -\frac{1}{2D_1^0}{\left  [-T_m^n -2W_{2m}^{2n} +
(2m+1)D_{m+1}^{n}
+\sum_{i=1}^{n-1}\sum_{j=1}^{i+1}D_j^iD_{m+1-j}^{n-i}
+\sum_{i=0}^{n-1}\sum_{j=0}^{i+1}C_j^iC_{m-j}^{n-i-1} \right   ]}
\end{eqnarray}
\begin{eqnarray}
C_{m}^{n} = -\frac{1}{2D_1^0}{\left  [-S_m^n -2W_{2m+1}^{2n+1} +
2(m+1)C_{m+1}^{n}
+2\sum_{i=1}^{n}\sum_{j=1}^{i+1}D_j^iC_{m+1-j}^{n-i} \right ]}
\end{eqnarray}
\begin{eqnarray}
E^{(n-1)} = \frac{1}{2}\left [T_0^n -D_1^n + 2W_{0}^{2n} -
\sum_{i=0}^{n-1}C_0^{i}C_0^{n-i-1}\right ]
\end{eqnarray}
Successively for the second excited state it is found that
\begin{eqnarray}
\left (x^2y^2 + Cxy + B\right )\left (\ddot{U}(x) +
\dot{U}(x)^2 -
2W(x) + 2E\right )
+ 2y^2 + 2\left (2xy^2 + yC\right )\dot{U}(x) = 0
\end{eqnarray}
Note that there is a misprint in \cite{stse}, in that $2x^2y$ should read $2xy^2$.
Once again
\begin{eqnarray}
\left [x^2y^2 + xy\sum_{n=1}^{\infty} c_n y^{2n} + \sum_{n=1}^{\infty} b_n
y^{2n}\right ]
\left[\sum_{n=0}^{\infty}\sum_{m=0}^{n}\left [T_m^nx^{2m}\right ]y^{2n}
+ \sum_{n=0}^{\infty}\sum_{m=0}^{n}\left [ S_m^nx^{2m+1}\right
]y^{2n+1}\right ]
\nonumber\\
+2y^2 + 2\left[2xy^2+y\sum_{n=1}^{\infty} c_n y^{2n}\right ]
\left [\sum_{n=0}^\infty \sum_{m=0}^{n+1} \left [{D_m^n}
x^{2m-1}\right ]y^{2n}
+\sum_{n=0}^{\infty} \sum_{m=0}^{n+1}\left  [
{C_m^nx^{2m}}\right   ] y^{2n+1}\right ] = 0
\end{eqnarray}
Now
\begin{eqnarray}
T_m^n = - \sum_{k=1}^{n-m} \left [ 2 c_k C_{m+1}^{n-k} + b_k
T_{m+1}^{n-k+1}
+c_k S_m^{n-k} \right ] - 4 D_{m+1}^n
\end{eqnarray}
\begin{eqnarray}
S_m^n = - \sum_{k=1}^{n-m} \left [ 2 c_k D_{m+2}^{n-k+1} + b_k
S_{m+1}^{n-k+1}
+c_k T_{m+1}^{n-k+1} \right ] - 4 C_{m+1}^n
\end{eqnarray}
\begin{eqnarray}
b_n = -\frac{1}{T_0^0}\left [{\sum_{k=1}^{n-1}
\left [ 2 c_k C_0^{n-k-1} + b_k
T_0^{n-k} \right ] + 2\delta_{n,1}}\right ]
\end{eqnarray}
\begin{eqnarray}
c_n = -\frac{1}{T_0^0 + 2D_1^0}\left [{\sum_{k=1}^{n-1} \left [2 D^{n-k}_1 +
T_0^{n-k}\right ] c_k+
\sum_{k=1}^{n} b_k S^{n-k}_0 +4C_0^{n-1}} \right ]
\end{eqnarray}
Employing the Taylor expansion technique the sign
of the $D_1^0$ coefficient has to be decided.
The two signs lead to distinct dissimilar solutions.
The choice leading to finite energy eigenvalues is
the negative solution, and this is the one we must choose.

\section{The results}
Exploying a computer algebraic algorithm programmed in MapleV \cite{maple},
all coefficients were calculated to a selected order of
approximation by recursion.

The energy estimates obtained using the $1/N$ method were
compared with strict analytical or accurate numerical results
from different sources \cite{stse,smb,Shi,ees,lar}.
In Table {\ref{table1}}, the results of the series
for the harmonic oscillator and the Coulomb potential
are displayed.
The approximation to the appropriate energy eigenvalues
is extremely good, indicating that in this case
the Taylor expansion procedure is exceedingly reliable.
The $1/N$ outcome for two additional potentials is
also shown in the Table {\ref{table1}}.
The results are excellent, but the energy sum starts
to diverge at some point in the approximation
(around 29th order or so), suggesting that the approximation
is correct to about 6-7 digits.
To illustrate the behaviour graphically, some plots of the
partial sums of the energy approximation versus the order of
approximation have been made for these potentials
(see figures \ref{fig1}, \ref{fig2} and \ref{fig3}).

Some potentials, also dealt with in \cite{stse,ees}, are
recomputed here to a much higher order of approximation
(see Tables \ref{table2}, \ref{table3} and \ref{table4}).
The results are very reasonable, and the high order of approximation
make the results nearly exact. In particular, it is observed that the
high-$l$ states approximate the correct energy quite well.
In Table {\ref{table4} for the 2nd excited states, only series
which did not begin to diverge to around 30th order of approximation
are presented.
Most of the approximated partial energy sums for the 1st and 2nd excited
states diverged at low $l$ in an oscillatory manner around the correct
energy eigenvalue.
A high-$l$ quantum number entirely cancelled
the divergence, leading to a seemingly convergent series.
As seen from the Tables {\ref{table3} and {\ref{table4} the
results for the harmonic oscillator and the Coulomb potential
are excellent and correspond exactly to the analytically known energy
eigenvalues.

The surveyed energy series are frequently seen to oscillate around
a certain energy value when plotted against the order of approximation.
The exact energy eigenvalue is found to be around this energy.
The series begins to diverge at
some order of approximation but still varies around the same definite
energy.
The oscillating series are often seen to fluctuate not termwise, but with
two terms or more above the specific energy value and then two terms
or more below, etc.

Some convention is required in order to discuss the results when
this behaviour is seen. In order to discover the best energy eigenvalue, the
last (first) term above and first (last) term below (or opposite)
the center of oscillation with least variation across it is found.
The result presented in the Tables is then these two terms.
For a normal oscillating series, the two terms with least variation across
the center of oscillation are presented.
In Tables {\ref{table2}}, {\ref{table3}} and {\ref{table4}} the
normal oscillating behaviour is observed.
The order of approximation of the partial sums employed is noted in the Tables.
For the convergent series, only the highest order term is presented.

To examine the oscillations more carefully another potential also discussed
in \cite{stse,ees}, namely the linear potential $2^{\frac{7}{2}}r$, will be analysed.
(the curious value of the constant is chosen such that the results in this paper can be
directly compared with those in \cite{stse,ees})
The results are presented in Table {\ref{table5}}.
For this potential the energy eigenvalue is calculated to
about the 29th order using the $1/N$ method for the first
three excited states.
The results show partial energy sums for the ground,
the 1st and 2nd excited state seemingly swinging around the correct
energy eigenvalue.
Some graphs showing the oscillations of the partial energy sums are
presented (see figures \ref{fig4}, \ref{fig5} and \ref{fig6}).

Different radial potentials for which ``exact'' results from \cite{ees}
are used, are shown in Tables {\ref{table6}}.
For most of these, satisfactory agreement with exact values is obtained.
Many of them are not termwise oscillating series.
The order of approximation employed gives a hint about the convergence of the
expansion.
The $r^{k}$ potentials where $k > 3$ are seen to diverge very quickly.
An empirical formulation of this principle would be ``the higher the power of the
potential, the poorer the convergence of the resulting $1/N$ energy series.''

In order to investigate the applicability of the $1/N$ procedure for potentials
with analytically precise eigenvalues, the method is now applied in the
case of fabricated potentials where the potential is set up from
the Schr\"odinger equation (see appendix \ref{appendix1}).
These examples illustrate the reliability of $1/N$ progressions
in the case of explicit, analytically known solutions and are hence
interesting for our purposes.
The results in Table {\ref{table7}} are obtained using potentials calculated
from a selected eigenfunction.
Most of the constructed potentials are seen in particular to have a not termwise
oscillatory behaviour.

Miscellaneous ``quark'' potentials and a particular double-well potential have
also been developed in $1/N$ series (see Tables {\ref{table8} and \ref{table9}}).
The double-well potential has also been handled in \cite{lar} using ordinary
perturbation calculations.

These results are not quite as good as those encountered for the other
potentials, but the Taylor method still yields a fairly good approximation to the
proper eigenvalue.

The double-well potential ($\frac{(r^2-R^2)^2}{8R^2}$, where $R$ is a constant)
is in fact a rather peculiar potential because
in a normal perturbation calculation one has to add powers of small terms
proportional to $\sim \exp(-\frac{2}{3}R^2)$ in order to take into account
quantum-mechanical tunnelling \cite{lar} in the perturbation series.
Such tunnelling considerations have not been included here, but
clearly for certain potentials such terms proportional to e.g. $\exp(-aN)$
should have an effect on the approximated energy eigenvalues. Maybe that is
why the result for the double-well potential deviates slightly from the
correct energy eigenvalue.
This should of course be investigated
in more details before using the Taylor expansion method more generally.
To demonstrate the behaviour of some of the approximations, plots of the
partial energy sums versus the order of convergence for some
of the potentials are presented (see figures \ref{fig7}, \ref{fig8},
\ref{fig9}, \ref{fig10}, \ref{fig11} and \ref{fig12}).

Most series are generated to about the 29th order of approximation with a numerical
precision of about 100 digits. In fact, it is feasible to generate the series
to the 100th (or higher) order of approximation with a numerical precision of about
1000 digits.
Further, it is conceivable to calculate the $1/N$-series exactly using
calculations with fractions and fractional roots to about the 100th order
of approximation.
The most important potentials are the harmonic oscillator,
the Coulomb potential, and with the constructed potentials examples where analytic
exact solutions are available.

Shanks resummed sums could have been used to resum the series. In a Shanks resummation
Shanks extrapolants are calculated from the partial sums, as
$$S_n = \frac{P_{n+1}\cdot P_{n-1} - P_n^2}{P_{n+1}+P_{n-1}-2P_n}$$
where $S_n$ is the $n$th Shanks extrapolant and $P_n$ the $n$th partial sum. See \cite{SP}.
In this paper it has been chosen not to do so because this question is an auxiliary aspect
of the technique.

\section{Discussion}
The major aspect of the long term behaviour of the $1/N$-series is of
course their apparently asymptotic behaviour.
As demostrated, many $1/N$ series match the exact energy satisfactorily when only
a certain number of terms are summed, but as further terms are added the sum
diverges oscillatingly to infinity. The plots presented clearly show
this behaviour with some minor differences.
Convergent $1/N$ series are also seen, such as the Coulomb and the harmonic
oscillator potentials series, together with the high-$l$ states 1/N series
(i.e. 1/k).
This behaviour is very similar to that of asymptotic series. Most perturbation
series are actually asymptotic.
An asymptotic series is, briefly stated, a series which begins to converge
towards a finite value, but which in the long run diverges.
Mathematically, an asymptotic power series in $\frac{1}{x}$
(here $x$ plays the rule of $N$) is a series for which
\begin{equation}
\lim_{|x| \rightarrow \infty} x^n\left [f(x) -
\sum_{r=0}^n\frac{a_r}{x^r}\right ] \rightarrow 0
\end{equation}
for all zero and positive $n$, (Poincar$\acute{\mathrm{e}}$'s definition)
\cite{as} or \cite{as2}
where $f(x)$ is a function and the sum is a partial sum for the asymptotic
power series of $f(x)$.

If the 1/$N$ series are asymptotic, the Coulomb and harmonic oscillator
potentials would then just be special cases in which the divergence is extremely
slow.
Comparing the eigenfunctions for the Coulomb and harmonic oscillator
potentials, they are both of the type $\exp(r^k)$, with $k=1$ and $k=2$,
respectively.
As a numerical experiment in this paper, eigenfunctions of the
type $\exp(r^k)$ are used to find the corresponding potential with
eigenvalue fixed at $E = 1$ (see appendix {\ref{appendix1}}).
These potentials are then expanded using the $1/N$ method,
(see Table {\ref{table7}}).
Examining this table one notices that the nearer $k$ is to 1 or 2, the
later the divergence of the resulting series, suggesting that the point of
divergence is related to $k$. For instance, with $k$ equal to 1.2 or 1.15,
the convergence is good to 25-30 orders or so.

One possible hint could be the radius of convergence for the Taylor
series of the $U(x)$ function, but further investigations examining the
$1/N$ series for, e.g., $U(x) = \cosh(x)$ have shown that there is no simple
relationship between the radius of convergence of
the Taylor series of the $U(x)$ function and the point of divergence of
its corresponding $1/N$ series. Furthermore, for $k=3$ the resulting series
is divergent, like the series for $k$ between 1 and 2.
In a more comprehensive treatment this should be inspected more carefully, and
the apparent relation between $k$ and the order of convergence of the
resulting $1/N$ series surveyed.

The series reported in this article are derived to a very high
numerical accuracy.
Strong evidence that numerical round-off errors have not affected the
results have been presented. By calculating results for
different power-law potentials it has been found that results obtained
``exactly''
(by fractions and fractional roots) and results obtained with an exactness
of 1000 digits and 100 digits, respectively, are nearly identical, and
that differences occur only at about the 60th-70th digit.

The purpose of this article has been to consider the convergence
of quantum mechanical $1/N$ series in detail. Many of the energy
series presented are extremely close to the exact energy eigenvalues.
The results clearly demonstrate that $1/N$ methods can be used
to obtain approximate energy eigenvalues
for different physical and chemical potentials.

The method has proved to yield fine results not only for the ground state
but also for the 1st and 2nd excited states. For the harmonic oscillator
and Coulomb potentials the results are extremely accurate.

The question of the convergence of the series has been
settled, and much evidence points towards the $1/N$ series being asymptotic.
Some of the series are so rapidly convergent at low orders that it will be
very difficult to see that they diverge at all.

It has been established that numerical round-off errors are
completely insignificant here. This is a very important
point because it demonstrates that the divergence problem is a feature
of the method itself and not a product of the computer arithmetic calculations.

\section{Acknowledgements}
This article could never have been written had it not been for Poul Henrik
Damgaard's help and advice. I also which to thank Patrick Ismail Ipsen for his
help and for discussions.

\newpage
\begin{appendix}
\section{How to construct the potential from a given ground-state
wavefunction}
\label{appendix1}
If $\phi(r) = \exp(\psi(r))$ is an eigenfunction of
\begin{equation}
\left  [-\frac{1}{2}\frac{d^2}{d r^2}-
\frac{1}{2}\frac{N-1}{r}\frac{d}{d r}
+ V(r) \right ]\phi(r) = E \phi(r)\label{Eqn:scr1}
\end{equation}
with eigenvalue $E$, then the potential must be
\begin{equation}
V(r) = -\frac12\frac{\partial^2 \psi(r)}{\partial r^2}+
\frac12\frac{\partial \psi(r)}{\partial r}^2+
\frac{1}{r}\frac{\partial \psi(r)}{\partial r} + E
\end{equation}
\section{The Schr\"{o}dinger equation in {\boldmath $N$} dimensions}
\label{appendix2}
In this appendix we shall see how to
derive the most important equation in
this paper, namely the radial Schr\"{o}dinger
equation in $N$ dimensions (see also \cite{louck}).

The Laplace operator in polar coordinates can be written \cite{louck} as
\begin{equation}
\nabla^2 = \frac{1}{h} \sum^{N-1}_{i=0} \frac{\partial}{\partial
\theta_i}\left ( \frac{h}{h^2_i} \frac{\partial}{\partial
\theta_i} \right  ) \label{Eqn:Laplace}
\end{equation}
where $\theta_0=r, h=\prod \nolimits^{N-1}_{j=0} h_j$ and
$h^2_i=\sum\nolimits^N_{j=1}\left (\frac{\partial x_j}{\partial
\theta_i}\right  )^2$.\\
The expressions for the spatial coordinates in $N$ dimensions are given by
\begin{eqnarray}
x_1 &=&
r\cos{\theta_1}\sin{\theta_2}\sin{\theta_3}\cdots\sin{\theta_{N
-1}}\nonumber \\
x_2 &=&
r\sin{\theta_1}\sin{\theta_2}\sin{\theta_3}\cdots\sin{\theta_{N
-1}}\nonumber \\
\vdots \nonumber \\
x_i &=&
r\cos{\theta_{i-1}}\sin{\theta_i}\sin{\theta_{i+1}}\cdots\sin{
\theta_{N-1}} \nonumber \\
\vdots \nonumber \\
x_N &=& r\cos{\theta_{N-1}}\label{Eqn:Coordinate}
\end{eqnarray}
for $N \geq 3$, $\left (x_1=r\cos{\theta_1}, \ x_2=r\sin{\theta_1};
N=2\right )$ and where $0\leq {r} < \infty$;
$0\leq {\theta_1}\leq 2\pi $;~$0\leq {\theta_i}\leq \pi $; $ 2\leq {i}\leq
N-1$ and $r^2=\sum\nolimits^N_{i=1} x^2_i$.\\
The functions $h_j$ are given by the following expressions
\begin{eqnarray}
h_0 &=& 1 \nonumber \\
h_1 &=&
r\sin{\theta_2}\sin{\theta_3}\cdots\sin{\theta_{N-1}}\nonumber \\
\vdots \nonumber \\
h_j &=&
r\sin{\theta_{j+1}}\sin{\theta_{j+2}}\cdots\sin{\theta_{N-1}}
\nonumber \\
\vdots \nonumber \\
h_{N-1} &=& r \label{Eqn:H}
\end{eqnarray}
Inserting the expression for $h_j$ in equation (\ref{Eqn:Laplace})
leads to
\begin{eqnarray}
\nabla^2 &=& \frac{1}{r^{N-1}}\frac{\partial}{\partial r} r^{N-1}
\frac{\partial}{\partial r}
+ \frac{1}{r^2}\sum^{N-2}_{i=1}\frac{1}{\prod^{N-1}_{j=i+1}\sin^2
{\theta_j}}
\left [\frac{1}{\sin^{i-1}{\theta_i}}\frac{\partial}{\partial
\theta_i}\sin^{i-1}{\theta_i}\frac{\partial}{\partial
\theta_i}\right  ] \nonumber \\
& & +
\frac{1}{r^2}\left [\frac{1}{\sin^{N-2}{\theta_{N-1}}}\frac{\partial}{\partial
\theta_{N-1}}\sin^{N-2}{\theta_{N-1}}\frac{\partial}{\partial
\theta_{N-1}}\right  ] \label{Eqn:Laplace2}
\end{eqnarray}
In general, the Hamiltonian can be written ($m = 1 = \hbar$)
\begin{equation}
\hat H = -\frac{1}{2}\nabla^2 = \frac{\hat p_r^2}{2} + \frac{\hat L^2}{2r^2}
\label{Eqn:Laplace0}
\end{equation}
separating angular momentum and radial momentum. Hence, from equation
(\ref{Eqn:Laplace0})
and equation (\ref{Eqn:Laplace2}) one can define total
angular momentum operators as
\begin{eqnarray}
\hat{L}^2_1 &=& - \frac{\partial ^2}{\partial
\theta^2_1}\nonumber \\
\hat{L}^2_2 &=& -
\left [\frac{1}{\sin{\theta_2}}\frac{\partial}{\partial
\theta_2}\sin{\theta_2}\frac{\partial}{\partial \theta_2}
-\frac{\hat{L}^2_1}{\sin^2{\theta_2}}\right  ]\nonumber \\
\vdots \nonumber \\
\hat{L}^2_i &=& -
\left [\frac{1}{\sin^{i-1}{\theta_i}}\frac{\partial}{\partial
\theta_i}\sin^{i-1}{\theta_i}\frac{\partial}{\partial \theta_i}
-\frac{\hat{L}^2_{i-1}}{\sin^2{\theta_i}}\right  ]
\nonumber \\
\vdots \nonumber \\
\hat{L}^2_{N-1} &=& -
\left [\frac{1}{\sin^{N-2}{\theta_{N-1}}}\frac{\partial}{\partial
\theta_{N-1}}\sin^{N-2}{\theta_{N-1}}\frac{\partial}{\partial
\theta_{N-1}}
-\frac{\hat{L}^2_{N-2}}{\sin^2{\theta_{N-1}}}\right  ]
\label{Eqn:Orbital}
\end{eqnarray}
In three dimensions, $\hat{L}^2_1$ is simply $\hat{L}^2_z$, and
$\hat{L}^2_2$ is $\hat{L}^2$. Also, it is seen that $\hat{L}^2_i$ has
the same expression independently of the number of spatial dimensions.

Equation (\ref{Eqn:Laplace2}) may now be recast as
\begin{equation}
\nabla^2 = \frac{1}{r^{N-1}}\frac{\partial}{\partial r} r^{N-1}
\frac{\partial}{\partial r} - \frac{\hat{L}^2_{N-1}}{r^2}
\end{equation}
The eigenvalues of the operator $\hat{L}^2_{N-1}$ will subsequently be
determined.

Only total angular momentum operators have so far been addressed, but
as in three dimensions, angular momentum components also exist.
They are written as
\begin{equation}
\hat{L}_{ij} = -\hat{L}_{ji} = \hat{x}_i\hat{p}_j -
\hat{x}_j\hat{p}_i \hspace{1cm} 1\leq i \leq j-1; 2\leq j \leq N
\end{equation}
where $\hat{p}_j$ is defined as
\begin{equation}
\hat{p}_j = -i \sum^{N-1}_{k=0} \left (
\frac{1}{h^2_k}\frac{\partial \hat{x}_j}{\partial
\theta_k}\right ) \frac{\partial}{\partial \theta_k}
\label{Eqn:Impuls}
\end{equation}
Considering the potential combinations, there are $\frac{N(N-1)}{2}$
distinct (not just differing by a sign) operators of the
pattern $\hat{L}_{ij}$.
The generalized angular momentum operators can then be obtained in
polar coordinates from equations (\ref{Eqn:Coordinate}),
(\ref{Eqn:Impuls}) and (\ref{Eqn:H}).
The results, i.e. the operators $\hat{L}_{ij}$, are then as follows
\begin{eqnarray}
\hat{L}_{12} &=& -i \frac{\partial}{\partial \theta_1}\nonumber
\\
\hat{L}_{13} &=& -i
\left  [\sin{\theta_1}\cot{\theta_2}\frac{\partial}{\partial
\theta_1} - \cos{\theta_1}\frac{\partial}{\partial
\theta_2}\right  ] \nonumber \\
\hat{L}_{23} &=& +i
\left  [\cos{\theta_1}\cot{\theta_2}\frac{\partial}{\partial
\theta_1} + \sin{\theta_1}\frac{\partial}{\partial
\theta_2}\right ] \nonumber \\
\hat{L}_{1j} &=& -i
\left [ \frac{\sin{\theta_1}\cot{\theta_{j-1}}}{\prod^{j-2}_{k=2}
\sin{\theta_k}}\frac{\partial}{\partial \theta_1} -
\sum^{j-2}_{k=2}\cot{\theta_k}\cos{\theta_1}\cos{\theta_{j-1}}
\left (
\frac{\prod^k_{h=2}\sin{\theta_h}}{\prod^{j-1}_{h=k+1}\sin{
\theta_h}}
\right )
\frac{\partial}{\partial \theta_k} \right ]\nonumber
\\ &+& i\left [\cos{\theta_1}\prod^{j-2}_{k=2}
\sin{\theta_k}\frac{\partial}{
\partial \theta_{j-1}}
\right ], 4\leq {j}\leq N\nonumber \\
\hat{L}_{2j} &=& +i
\left [\frac{\cos{\theta_1}\cot{\theta_{j-1}}}{\prod^{j-2}_{k=2}
\sin{\theta_k}}\frac{\partial}{\partial \theta_1} +
\sum^{j-2}_{k=2}\cot{\theta_k}\sin{\theta_1}\cos{\theta_{j-1}}
\left (
\frac{\prod^k_{h=2}\sin{\theta_h}}{\prod^{j-1}_{h=k+1}\sin{
\theta_h}}
\right  )
\frac{\partial}{\partial \theta_k} \right ]\nonumber \\ &+&
i\left [\sin{\theta_1}\prod^{j-2}_{k=2}\sin{\theta_k}\frac{\partial}{
\partial \theta_{j-1}} \right ],4\leq {j}\leq N \nonumber \\
\hat{L}_{ij} &=& -i
\left [\frac{\sin{\theta_{i-1}}\cot{\theta_{j-1}}}{\prod^{j-2}_{k=i}
\sin{\theta_k}}\frac{\partial}{\partial\theta_{i-1}} -
\sum^{j-2}_{k=i}\cot{\theta_k}\cos{\theta_{i-1}}\cos{\theta_{j-1}}
\left  (
\frac{\prod^k_{h=i}\sin{\theta_h}}{\prod^{j-1}_{h=k+1}\sin{\theta_h}}
\right   )
\frac{\partial}{\partial \theta_k} \right ] \nonumber \\
&+& i\left [\cos{\theta_{i-1}}\prod^{j-2}_{k=i}\sin{\theta_k}\frac{\partial}
{\partial \theta_{j-1}}
\right   ], 3\leq {i}\leq j-2 , \ 5\leq {j}\leq N\nonumber \\
\hat{L}_{j-1j} &=& -i
\left [\sin{\theta_{j-2}}\cot{\theta_{j-1}}\frac{\partial}
{\partial \theta_{j-2}} - \cos{\theta_{j-2}}\frac{\partial}{\partial
\theta_{j-1}}\right  ] ,\hspace{0.25cm} 4\leq {j}\leq N
\end{eqnarray}
for $N \geq 3, \left (h_0=1 , h_1=r ; N=2 \right )$ \cite{louck}.
For $N=3$, it is noted that $\hat{L}_{12}$ is simply $\hat{L}_z$,
$\hat{L}_{13}$ is $-\hat{L}_y$ and that $\hat{L}_{23}$ is $\hat{L}_x$.
As equation (\ref{Eqn:Orbital}) shows $\hat{L}^2_1,
\hat{L}^2_2,\ldots,\hat{L}^2_{N-1}$ clearly commute because they
depend on distinct angles, meaning that they have simultaneous
eigenfunctions. Since $\hat{L}^2_i$ is a sum of squares of
Hermitian operators of the form
($\hat x_i\hat p_j - \hat x_j \hat p_i$),
it is known that the eigenvalues of $\hat{L}^2_i$ will be real
and non-negative.
Furthermore, the eigenfunctions are orthogonal because they are
eigenfunctions of Hermitian operators. Assuming a proper normalisation,
they may be written as
\begin{equation}
Y\left (\lambda_{N-1},\ldots,\lambda_1\right  ) =
Y_{\lambda_{N-1},\ldots,\lambda_1}\left (\theta_1,\ldots,\theta_
{N-1}\right  ) =
\prod^{N-1}_{i=1}\Theta_i\left (\lambda_i,\lambda_{i-1}\right  )
\label{Eqn:Kuglef}
\end{equation}
where $\Theta_i\left (\lambda_i,\lambda_{i-1}\right  )$ is
a function only of $\theta_i$, and
$\Theta_1\left (\lambda_1,\lambda_0\right  )\equiv \Theta_1\left (\lambda
_1\right  )$ and
$\lambda_i$ is the associated eigenvalue of $\hat{L}^2_i$.\\
Clearly,
$\Theta_i\left (\lambda_i,\lambda_{i-1}\right  )$ will satisfy
\begin{eqnarray}
\hat{L}^2_1\Theta_1\left (\lambda_1\right  ) =
\lambda_1\Theta_1\left (\lambda_1\right  )\nonumber \\
\hat{L}^2_i\left (\lambda_{i-1}\right  )\Theta_i\left (\lambda_i,
\lambda_{i-1}\right  ) =
\lambda_i\Theta_i\left (\lambda_i,\lambda_{i-1}\right  ) \nonumber \\
\left (\hat{L}^2_i\left (\lambda_{i-1}\right  )-\hat{L}^2_{i-1}\left
(\lambda_{i-2}\right  )\right  )\Theta_{i-1}\left (\lambda_{i-1},
\lambda_{i-2}\right  )\Theta_i\left (\lambda_i,\lambda_{i-1}\right  )\nonumber
\\ =
\left (\lambda_i -
\lambda_{i-1}\right  )\Theta_{i-1}\left (\lambda_{i-1},
\lambda_{i-2}\right  )\Theta_i\left (\lambda_i,\lambda_{i-1}\right  )
\end{eqnarray}
for $2\leq {i}\leq N-1$.\\
Since
\begin{equation}
\hat L_i^2 \left (\lambda_{i-1} \right  ) - \hat L_{i-1}^2\left (\lambda_{i-2}\right  ) =
\sum_{j=1}^i \left  (\hat x_j\hat p_{i+1} - \hat x_{i+1}\hat p_j\right  )^2
\end{equation}
the eigenvalues of $\left (\hat L_i^2 \left (\lambda_{i-1} \right  )
- \hat L_{i-1}^2 \left (\lambda_{i-2} \right  ) \right  )  = \lambda_i - \lambda_{i-1}$
must be non-negative and real simply because the sum is the diagonal
matrix elements of the Hermitian
operator $(\hat x_j \hat p_{i+1} - \hat x_{i+1}\hat p_j)^2$ \cite{louck}.
Clearly
\begin{equation}
\lambda_i \geq \lambda_{i-1} \label{Eqn:Eigenv0}
\end{equation}
which implies that
\begin{equation}
\lambda_{N-1}\geq{\lambda_{N-2}}\geq\cdots\geq{\lambda_1}\geq{0}
\label{Eqn:Eigenv}
\end{equation}
because $\lambda_1 = m^2 \geq 0$\\
At $N=3$, $\lambda_1=l^2_1=m^2$ and $\lambda_2=l_2 \left
(l_2+1\right)=l \left (l+1 \right  ) = l\left (l + 2 - 1\right ) $\\
This suggests that in $(i-1)$ dimensions
\begin{equation}
\lambda_i = l_i \left ( l_i +i -1 \right  )
\end{equation}
Although the value of $l_i$ is unknown, a sensible guess would be
$l_i \in \mathbb{N}_0$ since it should be legitimate for $N=3$. By
the principle of induction it is feasible to prove this
proposition by proving that if
\begin{equation}
\hat{L}^2_{i-1} \left (l_{i-2}\right  ) \Theta_{i-1} \left (l_{i-1},l_{i-2}\right  )
= l_{i-1} \left (l_{i-1} +i-2\right  ) \Theta_{i-1} \left ( l_{i-1},l_{i-2} \right  )
\label{Eqn:Bevis}
\end{equation}
where $l_{i-1} \in \mathbb{N}_0$,\\ then
\begin{equation}
\hat{L}^2_i \left (l_{i-1}\right  ) \Theta_i
\left (l_i,l_{i-1}\right  ) = l_i \left (l_i +i
-1\right  ) \Theta_i \left (l_i,l_{i-1}\right )
\end{equation}
$l_{i} \in \mathbb{N}_0$,
and for a given $l_i = 0, 1, 2, ..$ one has $l_{i-1} = 0, 1, 2,..,l_{i}$.\\
{\sc Proof:}\\
Using (\ref{Eqn:Orbital}), $\hat L_i^2(l_{i-1})$ may be transformed into
\begin{equation}
\hat{L}^2_i\left (l_{i-1}\right  ) = -
\left [\frac{\partial^2}{\partial \theta^2_i}+
\left (i-1\right  )\cot{\theta_i}\frac{\partial}{\partial \theta_i}
-\frac{l_{i-1}\left (l_{i-1} +i
-2\right  )}{\sin^2{\theta_i}}\right  ]
\end{equation}
If ladder operators are defined as
\begin{eqnarray}
\hat{L}^+_i\left (l_{i-1}\right  ) &=& \frac{\partial}{\partial
\theta_i}-l_{i-1}\cot{\theta_i}\nonumber \\
\hat{L}^-_i\left (l_{i-1}\right  ) &=& -\frac{\partial}{\partial
\theta_i}-\left (l_{i-1}+i-2\right  )\cot{\theta_i}
\end{eqnarray}
then $\hat{L}^2_i\left (l_{i-1}\right  )$ can be written as
\begin{eqnarray}
\hat{L}^2_i\left (l_{i-1}\right  ) &=&
\hat{L}^+_i\left (l_{i-1}-1\right  )\hat{L}^-_i\left (l_{i-1}\right  ) +
\left (l_{i-1}+i-2\right  )\left (l_{i-1}-1\right  )\nonumber \\
&=& \hat{L}^-_i\left (l_{i-1}+1\right  )\hat{L}^+_i\left (l_{i-1}\right  ) +
l_{i-1}\left (l_{i-1}+i-1\right  ) \label{Eqn:L2+-}
\end{eqnarray}
By multiplying (\ref{Eqn:L2+-}) from right and left with
$\hat{L}^{\pm}_i\left (l_{i-1}\right  )$ one finds
\begin{eqnarray}
\hat{L}^2_i\left (l_{i-1}+1\right  )\hat{L}^+_i\left (l_{i-1}\right  ) &=&
\hat{L}^+_i\left (l_{i-1}\right  )\hat{L}^2_i\left (l_{i-1}\right  )\nonumber \\
\hat{L}^2_i\left (l_{i-1}-1\right  )\hat{L}^-_i\left (l_{i-1}\right  ) &=&
\hat{L}^-_i\left (l_{i-1}\right  )\hat{L}^2_i\left (l_{i-1}\right  )
\label{Eqn:L2L+-}
\end{eqnarray}
If now (\ref{Eqn:L2L+-}) operates on
$\Theta_i\left (\lambda_i,l_{i-1}\right  )$ it leads to
\begin{eqnarray}
\hat{L}^2_i\left (l_{i-1}+1\right  )\hat{L}^+_i\left (l_{i-1}\right  )\Theta_i\left
(\lambda_i,l_{i-1}\right  ) &=&
\lambda_i \hat{L}^+_i\left (l_{i-1}\right  )\Theta_i\left
(\lambda_i,l_{i-1}\right  ) \nonumber \\
&=& \lambda_i A\left (\lambda_i,l_{i-1}\right  )
\Theta_i\left (\lambda_i,l_{i-1}+1\right  ) \nonumber \\
\hat{L}^2_i\left (l_{i-1}-1\right  )\hat{L}^-_i\left (l_{i-1}\right  )\Theta_i\left
(\lambda_i,l_{i-1}\right  ) &=&
\lambda_i \hat{L}^-_i\left (l_{i-1}\right  )\Theta_i\left
(\lambda_i,l_{i-1}\right  ) \nonumber \\
&=& \lambda_i B\left (\lambda_i,l_{i-1}\right  )
\Theta_i\left (\lambda_i,l_{i-1}-1\right  )
\label{Eqn:L-B}
\end{eqnarray}
where $A,B$ are normalizing factors. It turns out that
$B\left (\lambda_i,l_{i-1}+1\right  )=A^\ast
 \left (\lambda_i,l_{i-1}\right  )$ \cite{louck}.\\
Thus, from (\ref{Eqn:L2+-})
\begin{eqnarray}
\hat{L}^2_i\left (l_{i-1} \right  )\Theta_i\left (\lambda_i,l_{i-1}\right  ) &=&
\lambda_i\Theta_i\left (\lambda_i,l_{i-1}\right  )\nonumber \\
&=& \left (|A\left (\lambda_i,l_{i-1}\right  )|^2 +
l_{i-1}\left (l_{i-1}+i-1\right  )\right  )\Theta_i\left
(\lambda_i,l_{i-1}\right  ) \nonumber \\
\Rightarrow \lambda_i - l_{i-1}\left (l_{i-1}+i-1\right  ) &=&
|A\left (\lambda_i,l_{i-1}\right  )|^2 \geq 0
\end{eqnarray}
As a consequence of (\ref{Eqn:Eigenv}), i.e. $\lambda_{i}\geq 0$,
there must exist a maximum value for $l_{i-1}$, say $l_i$, for a
given $\lambda_i$, such that
\begin{equation}
\hat{L}^+_i\left (l_i\right  )\Theta_i\left (\lambda_i,l_i\right  )=0
\end{equation}
where $l_{i}\in \mathbb{N}_0$, because $l_{i-1}\in \mathbb{N}_0$.
This leads to
\begin{equation}
\lambda_i=l_i\left (l_i +i -1\right  )
\end{equation}
where $l_i$ is a positive integer or zero.\\
To see what the possible values of $l_{i-1}$ are, given $l_{i}$,
one notes that clearly $l_{i-1} = l_i$ is plausible and that (\ref{Eqn:L-B})
implies that
\begin{equation}
L^-_i \left (l_{i}\right  ) \Theta_i \left (\lambda_i,l_i\right  )
\end{equation}
is an eigenfunction of $\hat{L}^2_i \left (l_{i}-1 \right  )$
with eigenvalue $\lambda_i$. Hence, $l_{i-1} = l_{i} - 1$ is viable.
Continuing this argument and introducing $l_{i-1}=l_i-1$
into (\ref{Eqn:L-B}), one discovers that $\hat{L}^2_i \left (l_i-2 \right  )$
has the eigenvalue $\lambda_i$. Consequently, another feasible
value for $l_{i-1}$ is $l_i-2$. Since 0 is the lowest
probable value of $l_{i-1}$ by presumption
and one diminish $l_{i-1}$ one by one, it is unquestionably seen that
the possible values of $l_{i-1}$ are
$l_i,l_i-1,l_i-2,..,0$ for $l_i \in \mathbb{N}_0$.\\
From (\ref{Eqn:Eigenv}) it is demonstrated that $\lambda_{N-1}$, and
accordingly $l_{N-1}$, have no upper boundaries.
It thus follows that $l_{N-1}=0,1,2,\ldots$, and in addition it is seen
that $l_2 \left (l_2+1 \right  ) \geq{l^2_1}\geq 0$, leading to
$l_1\leq |l_2|$.
Since (\ref{Eqn:Bevis}) is believed to be correct for $i=3$, it will also
be valid for $i=4$, and therefore by induction
\begin{eqnarray}
\hat{L}^2_i\Theta_i\left (l_i,l_{i-1}\right  ) &=& l_i\left (l_i +i
-1\right  )\Theta_i\left (l_i,l_{i-1}\right  ) \label{Eqn:Slut}
\end{eqnarray}
where $1\leq {i}\leq N-1$,
$0\leq {l_2}\leq {l_3}\leq {l_4}\leq\cdots\leq {l_{N-3}}\leq {l_{N-2}}\leq {
l_{N-1}}<\infty$
and $l_1\leq {|l_2|}$. \\
Equation (\ref{Eqn:Kuglef}) may now be written as
\begin{equation}
Y\left (l_{N-1},\ldots,l_1\right  )=\prod^{N-1}_{i=1}\Theta_i\left
(l_i,l_{i-1}\right  )
\end{equation}
and
\begin{equation}
\hat{L}^2_iY\left (l_{N-1},\ldots,l_i,\ldots,l_1\right  )=l_i\left
(l_i+i-1\right  )Y\left (l_{N-1},\ldots,l_i,\ldots,l_1\right  )
\end{equation}
Substituting $i = N-1$ into (8), one finally arrives at
\begin{equation}
\left  [-\frac{1}{2}\frac{d^2}{d r^2} -
\frac{1}{2}\frac{N-1}{r}\frac{d}{d r}
+ \frac{l(l + N - 2)}{2r^2} + V(r) \right   ]\phi(r) = E \phi(r)
\end{equation}
as the $N$-dimensional radial Schr\"odinger equation.
\end{appendix}

\newpage

\begin{figure}
\includegraphics*[width=8cm,angle=90]{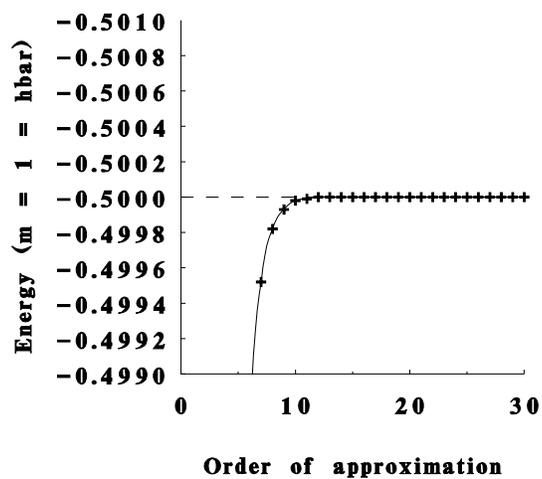}
\caption{Long-term behaviour of the $1/N$-energy
series for the Coulomb potential. The dotted line represents the
exact energy eigenvalue.}
\label{fig1}
\end{figure}

\begin{figure}
\includegraphics*[width=8cm,angle=90]{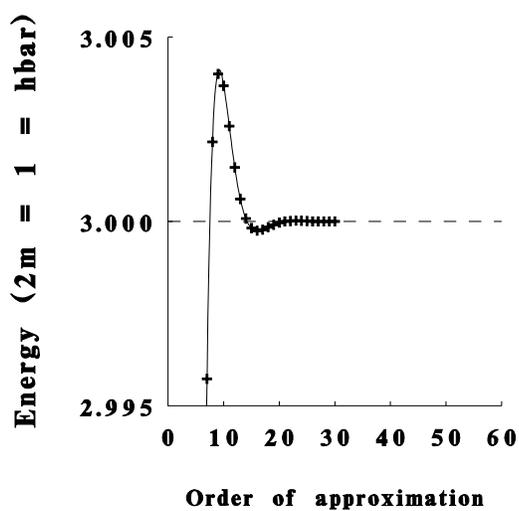}
\caption{Caption as for figure \ref{fig1} for
the harmonic oscillator potential.}
\label{fig2}
\end{figure}

\newpage

\begin{figure}
\includegraphics*[width=8cm,angle=90]{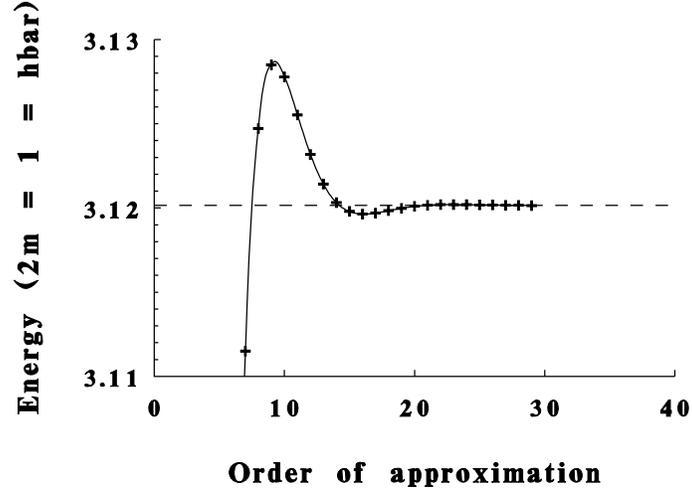}
\caption{Long-term behaviour for an analytic solvable
supersymmetic potential. Pot: $r^2+\frac{tr^2}{1+gr^2}$, $t=0.1$
and $g=0.1$. Angular momentum zero.\cite{Shi}} \label{fig3}
\end{figure}

\begin{figure}
\includegraphics*[width=8cm,angle=90]{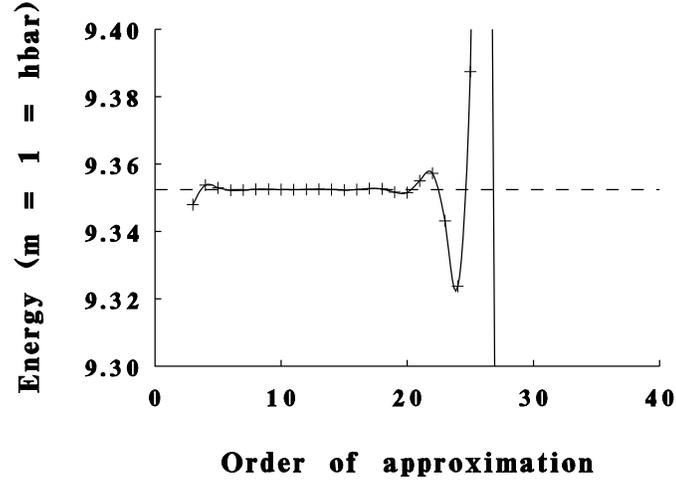}
\caption{Graph of the energy series for the ground state of the linear
potential: $2^{\frac{7}{2}}r$ with zero angular momentum: $l=0$.\\
It is seen that the convergence for the first 10-20 terms is excellent.
The series starts to diverge oscillatingly at about the 25th order of approximation.}
\label{fig4}
\end{figure}

\newpage

\begin{figure}
\includegraphics*[width=8cm,angle=90]{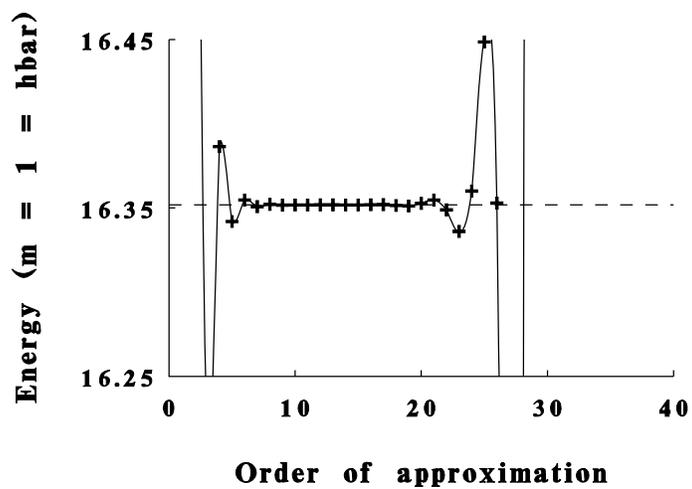}
\caption{Caption as for figure \ref{fig4} for the 1st excited state.}
\label{fig5}
\end{figure}

\begin{figure}
\includegraphics*[width=8cm,angle=90]{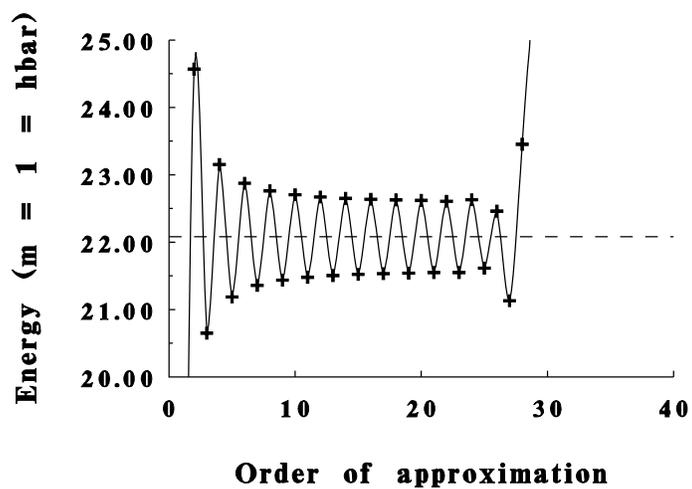}
\caption{Caption identical as for figure \ref{fig4} and \ref{fig5}
for the 2nd excited state.}
\label{fig6}
\end{figure}

\newpage

\begin{figure}
\includegraphics*[width=8cm,angle=90]{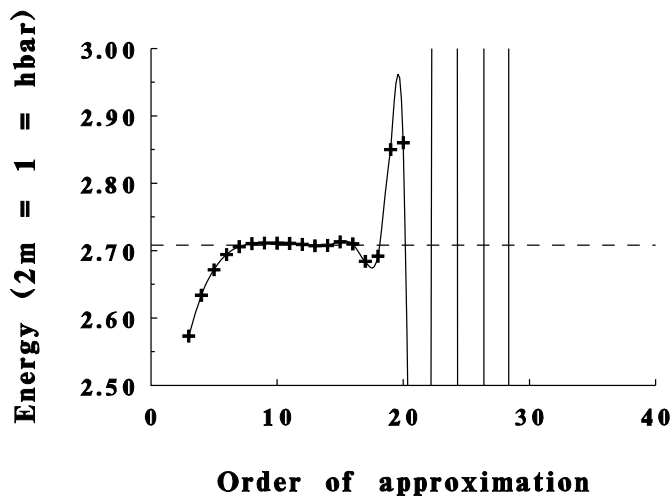}
\caption{Energy series for the ground state of the potential
$r^{\frac{3}{2}}$ with zero angular momentum as a function of the order
of approximation. As seen the convergence of the series is good up to a certain point, but
at about the 20th order of approximation it starts to diverge.
The series is seen to be oscillatory.}
\label{fig7}
\end{figure}

\begin{figure}
\includegraphics*[width=8cm,angle=90]{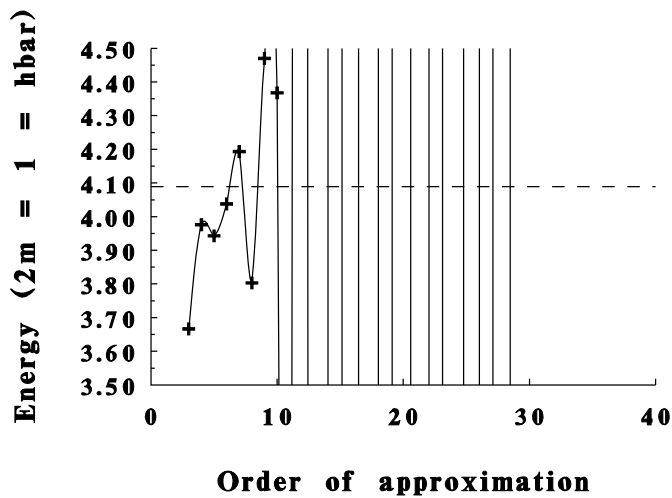}
\caption{Caption as for figure \ref{fig7} for the
potential $r^5$. Now the divergence is
much faster, and the series diverges even before it has reached its exact value
(the power effect).}
\label{fig8}
\end{figure}

\newpage

\begin{figure}
\includegraphics*[width=8cm,angle=90]{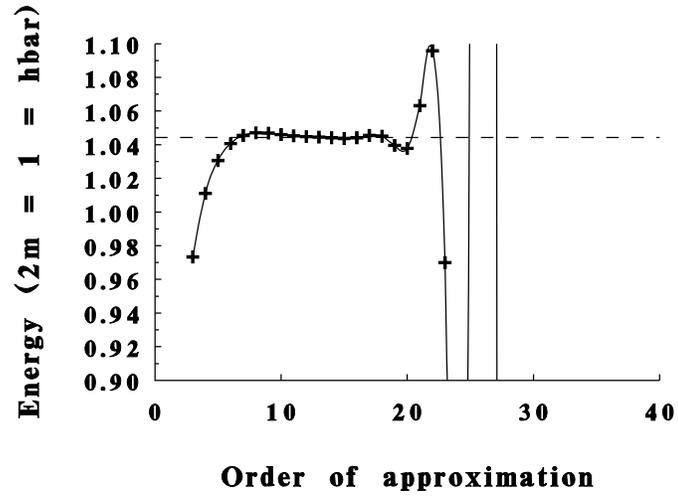}
\caption{Caption as for figure \ref{fig7} for the $\ln(r)$ potential.}
\label{fig9}
\end{figure}

\begin{figure}
\includegraphics*[width=8cm,angle=90]{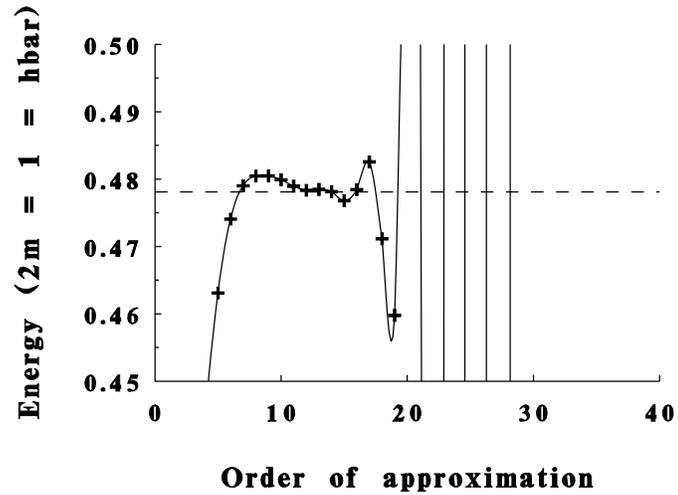}
\caption{The ground state of the quark potential
$\frac{r}{(2.34)^2} -\frac{52}{100r}$ with zero angular momentum is seen
to diverge at approximately the 20th order of approximation.}
\label{fig10}
\end{figure}

\newpage

\begin{figure}
\includegraphics*[width=8cm,angle=90]{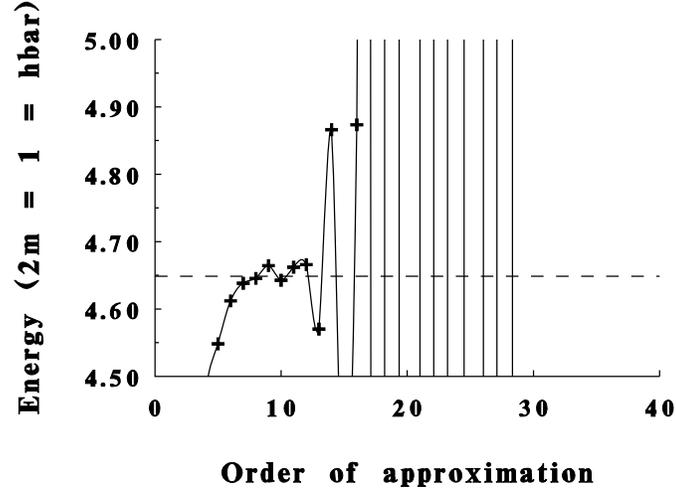}
\caption{Caption as for figure \ref{fig10} for the
quark potential $r^2+r^4$.
The divergence just starts much faster than in figure \ref{fig10}.}
\label{fig11}
\end{figure}

\begin{figure}
\includegraphics*[width=8cm,angle=90]{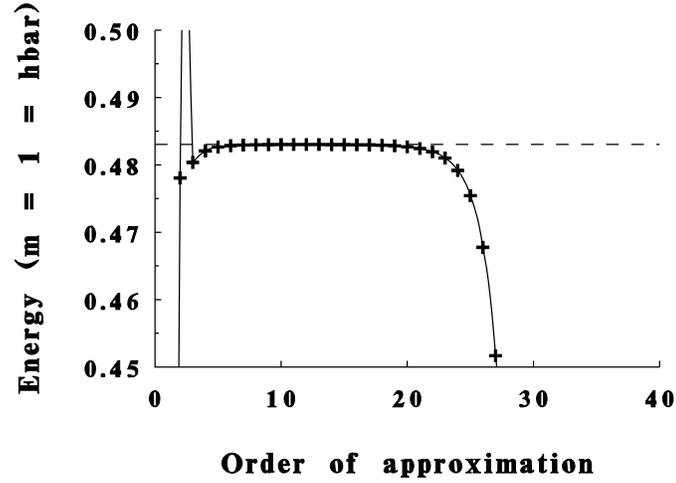}
\caption{Energy series for the ground state of the double-well potential
$\frac{(r^2-R^2)^2}{8R^2},\ R = 4$ with zero angular momentum.
It is seen that the $1/N$ series converges to about the 20th order
of approximation. It is also seen that oscillatory divergence does not
occur for this potential.}
\label{fig12}
\end{figure}
\newpage

\begin{table}
\caption{\label{table1} Table showing the convergence of the $1/N$
series for 4 different potentials ($2m = 1 = \hbar$), except for
the Coulomb potential in which $(m = 1 = \hbar)$. The exact energy
for the last potential is calculated in \cite{Shi} with a
supersymmetric method.}
\begin{ruledtabular}
\begin{tabular}{ccccc}
\multicolumn{4}{c}{\bf The partial sums of the energy approximation}\\
$r^2$ & $-\frac{1}{r}$ & $ r^2 + \frac{\frac{1}{10}\cdot r^2}{1+
\frac{1}{10}\cdot r^2}$ & $r^2 + \frac{\frac{406}{10000}
\cdot r^2}{1+\frac{1}{100}\cdot r^2}$ & Order of
approximation \\ \tableline
2.1213203433 & -0.22222222222 & 2.2154999 & 2.07828211 & 1 \\
2.9526299690 & -0.49108367627 & 3.0712441 & 2.89354432 & 5 \\
3.0036770689 & -0.49993508216 & 3.1238941 & 2.94360413 & 10 \\
2.9997496105 & -0.49999961670 & 3.1198959 & 2.93982608 & 15 \\
3.0000010566 & -0.49999999794 & 3.1200487 & 2.93996827 & 20 \\
2.9999982706 & -0.50000000000 & 3.1200752 & 2.93999831 & 29 \\
$3.0000000000$ & -0.50000000000 & diverges & diverges & 100 \\
\multicolumn{4}{c}{\bf Exact results for the eigenenergies}\\
$3_{\rm exact}$\cite{smb} & $-0.5_{\rm exact}$\cite{smb} &
$3.120081$ \cite{Shi} & $2.94_{\rm exact \ super \ sym.}$
\cite{Shi}
& $\sim$\\
\end{tabular}
\end{ruledtabular}
\end{table}

\begin{table}
\caption{\label{table2} Results for the ground state energy $(n =
0)$, $N =3$, for different potentials $(m = \hbar = 1)$. The
``exact'' results refer to \cite{stse,ees}; the calculated results
are referred to as $E_\mathrm{Taylor}$.}
\begin{ruledtabular}
\begin{tabular}{cccccc}
Potential & $l$ & $E_{\mathrm{shifted \ method}}$ \cite{ees}&
$E_{\mathrm{Taylor}}$ &
 $E_{\mathrm{exact_{num}}}$ \cite{stse,ees}  & Used order
\\ \tableline
$-2^{1.7}r^{-0.2}$ & 0 & -2.68601 & -2.685882 & -2.686 & 29 \\
$-2^{1.7}r^{-0.2}$ & 1 & -2.34494 & -2.344946 & -2.345 & 29\\
$-2^{1.7}r^{-0.2}$ & 2 & -2.15626 & -2.156260 & -2.156 & 29\\
$-2^{1.7}r^{-0.2}$ & 3 & -2.02906 & -2.029065 & -2.029 & 29\\
$-2^{0.8}r^{-0.8}$ & 0 & -1.21870 & -1.218693 & -1.218 & 29 \\
$-2^{0.8}r^{-0.8}$ & 1 & -0.50044 & -0.5004397 & -0.500 & 29\\
$-2^{0.8}r^{-0.8}$ & 2 & -0.29470 & -0.2946959 & -0.295 & 29\\
$-2^{0.8}r^{-0.8}$ & 3 & -0.20191 & -0.2019137 & -0.202 & 29\\
\end{tabular}
\end{ruledtabular}
\end{table}

\begin{table}
\caption{\label{table3} Results for the 1st excited state $(n =
1)$, $N = 3$, for different potentials $(m = \hbar = 1)$. The
``exact'' results refer to \cite{stse,smb,ees}; the calculated
results are referred to as $E_\mathrm{Taylor}$.}
\begin{ruledtabular}
\begin{tabular}{cccccc}
Potential & $l$ & $E_{\mathrm{shifted \ method}}$ \cite{ees} &
$E_{\mathrm{Taylor}}$
&$E_{\mathrm{exact_{num}}}$ \cite{stse,ees}
& Used order
\\ \tableline
$-2^{1.7}r^{-0.2}$ & 0 & -2.25483
& -2.253515 - (-2.25314) & -2.253 & 16-17\\
$-2^{1.7}r^{-0.2}$ & 1 & -2.10103 & -2.100738 & -2.101 & 29\\
$-2^{1.7}r^{-0.2}$ & 2 & -1.99015 & -1.990056 & -1.990 & 29\\
$-2^{1.7}r^{-0.2}$ & 3 & -1.90491 & -1.904867 & -1.905 & 29\\
$-2^{0.8}r^{-0.8}$ & 0 & -0.46282 & -0.462291 & -0.462 & 22\\
$-2^{0.8}r^{-0.8}$ & 1 & -0.28071 & -0.280648 & -0.281 & 29\\
$-2^{0.8}r^{-0.8}$ & 2 & -0.19493 & -0.194912 & -0.195 & 29\\
$-2^{0.8}r^{-0.8}$ & 3 & -0.14635 & -0.146342 & -0.146 & 29\\ \tableline
Potential & $l$ && $E_{\mathrm{Taylor}}$
& $E_{\mathrm{exact}}$ \cite{smb} & Used order \\ \tableline
$-\frac{1}{r}$ & 0 && -0.12500000 & -0.125$_{\mathrm{exact}}$ & 29\\
$-\frac{1}{r}$ & 1 && -0.05555556 & -0.0$\overline{5}_{\mathrm{exact}}$ & 29\\
$-\frac{1}{r}$ & 2 && -0.03125000 & -0.03125$_{\mathrm{exact}}$ &  29\\
$-\frac{1}{r}$ & 3 && -0.02000000 & -0.02$_{\mathrm{exact}}$ &  29\\
$-\frac{1}{r}$ & 4 && -0.01388889 & -0.013${\overline{8}_\mathrm{exact}}$ & 29\\
$\frac{1}{2}r^2$ & 0 && 3.50000000 & 3.5$_{\mathrm{exact}}$ & 29\\
$\frac{1}{2}r^2$ & 1 && 4.50000000 & 4.5$_{\mathrm{exact}}$ & 29\\
$\frac{1}{2}r^2$ & 2 && 5.50000000 & 5.5$_{\mathrm{exact}}$ & 29\\
$\frac{1}{2}r^2$ & 3 && 6.50000000 & 6.5$_{\mathrm{exact}}$ & 29\\
$\frac{1}{2}r^2$ & 4 && 7.50000000 & 7.5$_{\mathrm{exact}}$ & 29\\
\end{tabular}
\end{ruledtabular}
\end{table}

\begin{table}
\caption{\label{table4} Results for the 2nd excited state $(n =
2)$, $N = 3$, for different potentials $(m = \hbar = 1)$. The
``exact'' results refer to \cite{stse,smb,ees}; the calculated
results are referred to as $E_\mathrm{Taylor}$.}
\begin{ruledtabular}
\begin{tabular}{cccccc}
Potential & $l$ & $E_{\mathrm{shifted \ method}}$ \cite{ees} &
$E_{\mathrm{Taylor}}$
&
$E_{\mathrm{exact_{num}}}$ \cite{stse,ees} & Used order\\ \tableline
$-2^{1.7}r^{-0.2}$ & 1 & -1.95147 & -1.950722 & -1.951 & 29 \\
$-2^{1.7}r^{-0.2}$ & 2 & -1.87535 & -1.875032 & -1.875 & 29 \\
$-2^{1.7}r^{-0.2}$ & 3 & -1.81266 & -1.812502 & -1.812 & 29 \\
$-2^{0.8}r^{-0.8}$ & 1 & -0.18745 & -0.187318 - (-0.187320) & -0.187 & 28-29\\
$-2^{0.8}r^{-0.8}$ & 2 & -0.14202 & -0.141980 & -0.142 & 29\\
$-2^{0.8}r^{-0.8}$ & 3 & -0.11280 & -0.112788 & -0.113 & 29\\
\tableline Potential & $l$ && $E_{\mathrm{Taylor}}$ &
$E_{\mathrm{exact}}$ \cite{smb} & Used order \\ \tableline
$-\frac{1}{r}$ & 0 && diverges & -0.0$\overline{5}_{\mathrm{exact}}$ & 29\\
$-\frac{1}{r}$ & 1 && -0.03125000 & -0.03125$_{\mathrm{exact}}$ & 29\\
$-\frac{1}{r}$ & 2 && -0.02000000 & -0.02$_{\mathrm{exact}}$ &  29\\
$-\frac{1}{r}$ & 3 && -0.01388889 & -0.013$\overline{8}_{\mathrm{exact}}$ &  29\\
$-\frac{1}{r}$ & 4 && -0.01020408 &
$-\frac{1}{96}_{\mathrm{exact}} \approx
-0.01020408$ & 29\\
$\frac{1}{2}r^2$ & 0 && 5.50000000 & 5.5$_{\mathrm{exact}}$ & 29\\
$\frac{1}{2}r^2$ & 1 && 6.50000000 & 6.5$_{\mathrm{exact}}$ & 29\\
$\frac{1}{2}r^2$ & 2 && 7.50000000 & 7.5$_{\mathrm{exact}}$ & 29\\
$\frac{1}{2}r^2$ & 3 && 8.50000000 & 8.5$_{\mathrm{exact}}$ & 29\\
$\frac{1}{2}r^2$ & 4 && 9.50000000 & 9.5$_{\mathrm{exact}}$ & 29\\
\end{tabular}
\end{ruledtabular}
\end{table}

\begin{table}
\caption{\label{table5} Table showing partial energy sums for the
potential: $2^{\frac{7}{2}}r$. Results for the first three excited
states are shown ($m = 1 = \hbar$).}
\begin{ruledtabular}
\begin{tabular}{ccccc}
Ground-state energy & 1st excited state energy & 2nd excited state energy
& Order of approximation
\\ \tableline
9.35240 & 16.35180 & 22.70413 & 10\\
9.35229 & 16.35175 & 21.52097 & 15\\
9.35153 & 16.35274 & 22.62047 & 20\\
9.38745 & 16.44861 & 21.61422 & 25\\
9.54245 & 16.35286 & 22.45952 & 26\\
9.24170 & 15.68014 & 21.13008 & 27\\
7.95750 & 15.92972 & 23.45280 & 28\\
9.26333 & 21.51171 & 25.79375 & 29\\
\multicolumn{4}{c}{\bf Exact energies \cite{stse,ees}} \\
9.35243$_{\mathrm{exact}}$ & 16.3518$_{\mathrm{exact}}$ &
22.08224$_{\mathrm{exact}}$ & $\sim$ \\
\multicolumn{4}{c}{\bf Shifted method \cite{ees}} \\
9.35243 & 16.32636 & 22.02319 & $\sim$ \\
\end{tabular}
\end{ruledtabular}
\end{table}
\clearpage
\begin{table}
\caption{\label{table6} Results for the ground state energy $(l =
n = 0)$, $N=3$, for different potentials $(2m = \hbar = 1)$. For
the ``exact'' results, reference is made to \cite{ees};
$E_\mathrm{Taylor}$ refers to the calculated results.}
\begin{ruledtabular}
\begin{tabular}{ccccc}
Potential & $E_{\mathrm{shifted \ method}}$ \cite{ees}
& $E_{\mathrm{Taylor}}$
&
$E_{\mathrm{exact_{num}}}$ \cite{ees} & Used order
\\               \tableline
$-r^{-1.5}$ & -0.29888 & -0.29880 - (-0.29931)  & -0.29609 & 27-28\\
$-r^{-1.25}$ & -0.22035 & -0.22038 & -0.22029 & 29\\
$r^{0.15}$ & 1.32795 & 1.32781 - 1.32797 & 1.32795 & 15-16 \\
$r^{0.5}$ & 1.83341 & 1.83361 - 1.83287 & 1.83339 & 13-14 \\
$r^{0.75}$ & 2.10815 & 2.10769 - 2.11223 & 2.10814 & 15-16 \\
$r^{1.5}$ & 2.70806 & 2.70780 - 2.71299 & 2.70809 & 14-15 \\
$r^{3}$ & 3.45111 & 3.46173 - 3.43840 & 3.45056 & 11-12 \\
$r^{4}$ & 3.80139 & 3.81467 - 3.37768 & 3.79967 & 7-8 \\
$r^{5}$ & 4.09146 & 4.19331 - 4.03767 & 4.08916 & 6-7 \\
$\ln(r)$ & 1.04436 & 1.04457 - 1.04414 & 1.0443 & 13-14 \\
\end{tabular}
\end{ruledtabular}
\end{table}

\begin{table}
\caption{\label{table7} Results for the ground state energy with
$(l = n = 0)$ and $N=3$, for different constructed potentials with
$(2m = \hbar = 1)$. The exact energies are chosen to be 1;
$E_{\mathrm{Taylor}}$ refers to the calculated results.}
\begin{ruledtabular}
\begin{tabular}{ccc}
Eigenfunction & $E_{\mathrm{Taylor}}$ & Used order \\
\tableline
$\exp({-r^{0.80}})$ & 1.0005 - 0.99917 & 12-13\\
$\exp({-r^{0.85}})$ & 1.000005 - 0.999993 & 29-30 \\
$\exp({-r^{0.90}})$ & 1.00003 - 0.999998 & 23-24 \\
$\exp({-r^{0.95}})$ & 1.0000 - 0.99999 & 24-25 \\
$\exp({-r^{1.05}})$ & 1.0000 - 1.00000 & 24-25 \\
$\exp({-r^{1.10}})$ & 1.0000 - 0.99998 & 25-26 \\
$\exp({-r^{1.15}})$ & 1.0001 - 1.00000 & 18-19 \\
$\exp({-r^{1.20}})$ & 1.0000 - 0.99998 & 22-23 \\
$\exp({-r^{1.25}})$ & 1.0000 - 0.99995 & 18-19\\
$\exp({-r^{1.30}})$ & 1.0000 - 0.99991 & 18-19 \\
$\exp({-r^{1.35}})$ & 1.0001 - 0.99985 & 18-19 \\
$\exp({-r^{1.40}})$ & 1.0002 - 0.99975 & 18-19 \\
$\exp({-r^{1.45}})$ & 1.0004 - 0.99960 & 18-19 \\
$\exp({-r^{1.50}})$ & 1.0004 - 0.99953 & 12-13 \\
$\exp({-r^{1.55}})$ & 1.0005 - 0.99958 & 12-13 \\
$\exp({-r^{1.60}})$ & 1.0006 - 0.99960 & 12-13 \\
$\exp({-r^{1.65}})$ & 1.0008 - 0.99960 & 12-13 \\
$\exp({-r^{1.70}})$ & 1.0002 - 0.99996 & 15-16 \\
$\exp({-r^{1.75}})$ & 1.0005 - 0.99990 & 15-16 \\
$\exp({-r^{1.80}})$ & 1.0010 - 0.99973 & 15-16 \\
$\exp({-r^{1.85}})$ & 1.0015 - 0.99944 & 14-15 \\
$\exp({-r^{1.90}})$ & 1.0019 - 0.99963 & 14-15 \\
$\exp({-r^{1.95}})$ & 1.0016 - 0.99968 & 14-15 \\
\end{tabular}
\end{ruledtabular}
\end{table}

\begin{table}
\caption{\label{table8} Results for the ground-state energy $(l =
n = 0)$, $N = 3$, for different potentials $(2m = \hbar = 1)$. For
the ``exact'' results, reference is made to \cite{ees}, again
$E_{\mathrm{Taylor}}$ refers to the calculated results.}
\begin{ruledtabular}
\begin{tabular}{ccccc}
Potential & $E_{\mathrm{shifted \ method}}$ \cite{ees} &
$E_{\mathrm{Taylor}}$ & $E_{\mathrm{exact_{num}}} $\cite{ees} &
Used order
\\           \tableline
$6.8698r^{0.1}-8.064$ & -0.31914 & -0.31936 - (-0.31883) & -0.31917 & 11-10\\
$\frac{r}{(2.34)^2} - 0.52\frac{1}{r}$ & 0.47880 & 0.48049 - 0.47990 & 0.47811 & 10-9\\
$r^2+r^4$ & 4.65061 & 4.66212 - 4.64267 & 4.64881 & 11-10 \\
\end{tabular}
\end{ruledtabular}
\end{table}

\begin{table}
\caption{\label{table9} Double-well potential:
$\frac{(r^2-R^2)^2}{8R^2}$, $R=4$. Results for the ground-state
energy $(l = n = 0)$, $(m = 1 = \hbar)$, $N = 3$.}
\begin{ruledtabular}
\begin{tabular}{cccc}
$E_{\mathrm{num}}$ \cite{lar} & $E_{\mathrm{Taylor}}$ &
$E_{\mathrm{perturbation}}$ \cite{lar} & Used order\\ \tableline
0.483053433 & 0.483018 - 0.483015 & 0.483053390 & 13-12\\
\end{tabular}\\
\end{ruledtabular}
\end{table}

\end{document}